\documentstyle[aps,prd,twocolumn,epsf,epsfig]{revtex}

\newcommand \bea{\begin{eqnarray}}
\newcommand \eea{\end{eqnarray}}
\newcommand \ga{\raisebox{-.5ex}{$\stackrel{>}{\sim}$}}
\newcommand \la{\raisebox{-.5ex}{$\stackrel{<}{\sim}$}}
\newcommand{\av}[1]{\langle{#1}\rangle}

\begin{document}
\twocolumn[\hsize\textwidth\columnwidth\hsize
\csname@twocolumnfalse%
\endcsname
\draft
\title{Pairing of fermions in atomic traps and nuclei}
\author{H. Heiselberg}
\address{Univ. of Southern Denmark, Campusvej 55, DK-5230 Odense M, Denmark}
\maketitle

\begin{abstract}
Pairing gaps for fermionic atoms in harmonic oscillator traps are
calculated for a wide range of interaction strengths and particle
numbers, and compared to pairing in nuclei. Especially systems where
the pairing gap exceeds the single level spacing but is smaller than the
shell splitting $\hbar\omega$, are studied which applies to most
trapped Fermi atomic systems as well as to finite nuclei. When solving
the gap equation for a large trap with such multi-level pairing, one
finds that the matrix elements between nearby harmonic oscillator
levels and the quasi-particle energies lead to a double logarithm of
the gap, and a pronounced shell structure at magic numbers. It is
argued that neutron and proton pairing in nuclei belongs to the class
of multi-level pairing, that their shell structure follows naturally
and that the gaps scale as $\sim A^{-1/3}$ - all in qualitative
agreement with odd-even staggering of nuclear binding energies.
Pairing in large systems are related to that in the bulk limit.  For
large nuclei the neutron and proton superfluid gaps approach the
asymptotic value in infinite nuclear matter: $\Delta\simeq 1.1$~MeV.
\\ \pacs{PACS numbers: 03.75.Fi, 21.65.+f, 74.20.Fg, 67.60.-g, 05.30.Fk}
\end{abstract}
\vskip1pc]

\section{Introduction}

Pairing in Fermi systems is central for understanding
superconductivity in condensed matter physics \cite{BCS},
superfluidity and glitches in neutron stars, excitations spectra and
odd-even staggering of binding energies in nuclei \cite{Pines},
metallic clusters \cite{metalc} and superconducting grains
\cite{grains}. New insight into the general properties of such Fermi
systems can now be obtained from experiments with atomic gases
\cite{JILA} which recently have been cooled down to degenerate
temperatures around 10\% of the Fermi temperature
\cite{Thomas,Jin,Paris,Ketterle}. The interactions between trapped $^6$Li and
$^{40}$K atoms have simultaneously been tuned above a Feshbach
resonance, where they become strongly attractive, in order to produce
optimal conditions for superfluidity \cite{SHSH}.  Atomic gases have widely
tunable number of particles, densities, interaction strengths,
temperatures, spin states, and other parameters which holds great
promise for a more general understanding of pairing phenomena in
solids, metallic clusters, grains, nuclei and neutron stars.

In the experiments with $^6$Li atoms the anisotropic expansion after
sudden release from the trap is as predicted from hydrodynamics
\cite{Stringari}. This is compatible both with a strongly interacting
superfluid and collisional hydrodynamics. The scattering length is
attractive $a<0$ and large such that densities above $\rho\gg 1/|a|^3$
are achieved and found mechanically stable against collapse. The
interaction energy per particle, like Fermi energy, scales as
$\rho^{2/3}$ \cite{Thomas,Jin,Paris} as predicted theoretically
\cite{HH,Carlson}. The superfluid gaps in bulk are also expected to be
of order the Fermi energy.

In nuclear physics pairing is observed directly in the odd-even
staggering of binding energies, i.e., that nuclei with an even number
of protons or neutrons are more strongly bound than nuclei with
corresponding odd numbers by a pairing gap of order $\sim 1$~MeV.  The
same pairing gap also determines the excitation energies of ground
state nuclei. On average the pairing energies in nuclei decrease with
the number of protons and neutrons modulated with a distinct shell
structure, i.e., they are smaller near closed shells. It would be most
interesting if the pairing gaps of fermionic atoms in harmonic
oscillator traps would display similar shell structure and scaling
with the number of particles as nuclei.

 The aim of this paper is to calculate pairing gaps in ultracold
atomic Fermi gases in harmonic oscillator traps and in nuclei, which
are then compared to data on odd-even staggering of nuclear binding
energies. For the atomic traps existing pairing gap calculations
\cite{HM,BH} are extended to systems, where level spectra and shell
effects play an important role. It is shown that such pairing
mechanisms are similar in nuclei, which to first approximation can be
considered as a finite system of fermions in a harmonic oscillator
field with attractive delta function interactions. Another important
common feature is the anharmonic fields that lead to splitting of the
single particle states which reduce the pairing to the levels closest
to the Fermi surface and leads to a distinct shell structure.  For a
wide range of interaction strengths and number of particles the
trapped atomic clouds are predicted to display similar scaling and
shell structure as nuclei as also seen in the experimental data on
neutron and proton pairing. The scaling with particle number and the
continuum or BCS limits are calculated for both large traps and nuclei
and thereby the pairing gaps are estimated for nuclear matter, which
also gives an idea about the superfluid gaps in neutron star matter.

 The paper is organized as follows. In section II the basic properties
of interacting fermions in harmonic oscillator traps are given, in
particular the single particle level spectra. Pairing is treated in
section III with individual subsections devoted to each pairing regime
where the last three subsection outline the novel bulk, strongly
interacting and multi-level pairing regime. In section IV we turn to
nuclei and show that they belong to the class of multi-level pairing.

\section{Dilute Fermi gases in harmonic traps}

We treat a gas of $N$ fermionic atoms of mass $m$ in a 
harmonic oscillator (h.o.) potential at zero temperature interacting
via a two-body interaction with attractive s-wave scattering length $a<0$. 
We shall mainly discuss a spherically symmetric trap and a dilute gas
(i.e. where the density $\rho$ obeys the condition
$\rho|a|^3\ll 1$) of particles
with two spin states with equal population. The Hamiltonian is then given by
\bea  \label{H}
  H &=& \sum_{i=1}^{N} \left( \frac{{\bf p}_i^2}{2m} +
    \frac{1}{2} m\omega^2 {\bf r}_i^2 \right)
  + g \sum_{i<j} 
     \delta^3({\bf r}_i-{\bf r}_{j}) \,,
\eea 
with the effective coupling $g=4\pi\hbar^2a/m$.
For a large number of particles $N$ at zero temperature the Fermi energy is
for a non-interacting system
\bea
   E_F=(n_F+3/2)\hbar\omega \,
  \simeq \, \left(3N\right)^{1/3} \hbar\omega   \,, \label{EF}
\eea
where $n_F\simeq(3N)^{1/3} $ is the h.o. quantum number at the Fermi surface.
The h.o. shells are highly degenerate with states having angular momenta
$l=n_F,n_F-2,...,1$ or 0 due to the     
$U(3)$ symmetry of the 3D spherically symmetric h.o. potential.
However, interactions split this degeneracy.
In the Thomas-Fermi (TF) approximation (see, e.g., \cite{PS}) the 
mean-field potential is
\bea
    U(r) &=& \frac{1}{2}g\rho(r)= 2\pi\frac{\hbar^2a}{m} \rho(r) \,, \label{U}
\eea
the Fermi energy 
\bea
  E_F=\frac{\hbar^2k_F^2(r)}{2m}+\frac{1}{2}m\omega^2r^2+U(r) \, .
\eea
The density can be determined from the Kohn-Sham energy density functionals.
In a dilute gas the mean field is small as compared to the Fermi energy
and its effect on the density can ignored in the following. Thus
\bea
     \rho(r) = k_F^3(r)/3\pi^2 \,
     \simeq \rho_0 \left(1-r^2/R_{TF}^2\right)^{3/2} \,, \label{rho} 
\eea
inside the cloud $r\le R_{TF}=a_{osc}\sqrt{2n_F+3}$, where
$a_{osc}=\sqrt{\hbar/m\omega}$ is the oscillator length, and
$\rho_0=(2n_F+3)^{3/2}/3\pi^2a_{osc}^3$ the central density. \cite{TF3} 

The splitting of the h.o. shells
degenerate levels $l=n_F,n_F-2,...,1$ or 0 in
the shell $n_F$ by the mean-field potential 
can be calculated perturbatively in the dilute limit.
An excellent approximation for the radial h.o. wave function 
with angular momentum $l$ and $(n_F-l)/2$ radial nodes in the h.o. shell
when $n_F\gg1$ is the WKB one \cite{WKB}
\bea
   {\cal R}_{n_Fl}(r)\simeq \frac{2}{\sqrt{\pi}}
   \frac{\sin(k_lr+\theta)}{k_lr} \,, \label{RWKB}
\eea
between turning points \cite{WKB}.
The phase $\theta$ will not be important in the following. 
The WKB wave number $k_l(r)$ is given by
\bea \label{kl}
 k_l^2(r)=\frac{2n_F+3}{a^2_{osc}}-\frac{r^2}{a^4_{osc}}-\frac{l(l+1)}{r^2} \,.
\eea

The single particle energies are
\bea 
  \epsilon_{n_F,l}&-&\left(n_F+\frac{3}{2}\right)\hbar\omega \, = \int U(r) 
 |{\cal R}_{n_Fl}(r)|^2 r^2dr \nonumber\\
  &=& \frac{2}{3\pi}\frac{a}{a_{osc}} (2n_F+3)^{3/2} \hbar\omega\,
\left[\frac{4}{3\pi}-\frac{1}{4\pi} \frac{l(l+1)}{n_F^2}\right]
 \,. \label{EMF}
\eea
The latter result is calculated from the overlap between the mean field
as given by Eqs. (\ref{U}) and  (\ref{rho}) and the WKB wave-functions
as given by Eqs. (\ref{RWKB}) and  (\ref{kl}) \cite{HM}. 
It is exact within WKB 
for $l\ll n_F$ and it compares well to numerical results for all $l$.
For attractive interactions ($a<0$)
the lowest-lying states have small angular momentum, which is {\it opposite}
to nuclei.

An important quantity for pairing is the {\it supergap} 
which was introduced in
Ref. \cite{HM} and will be discussed below 
\bea \label{G}
 G = \frac{32\sqrt{2n_F+3}}{15\pi^2} \frac{|a|}{a_{osc}}\hbar\omega \,.
\eea
In comparison the total shell splitting in a shell $n_F$ is from Eq.
(\ref{EMF})
\bea
  D \equiv |\epsilon_{n_Fl=n_F}-\epsilon_{n_Fl=0}| 
  \simeq \frac{5(n_F+3/2)}{32} G  . \label{D}
\eea

The pairing depends crucially on the shell splitting $D$ and the
supergap $G$ as will be shown in the following section.

\vspace{0.cm}
\begin{figure}
\begin{center}
\psfig{file=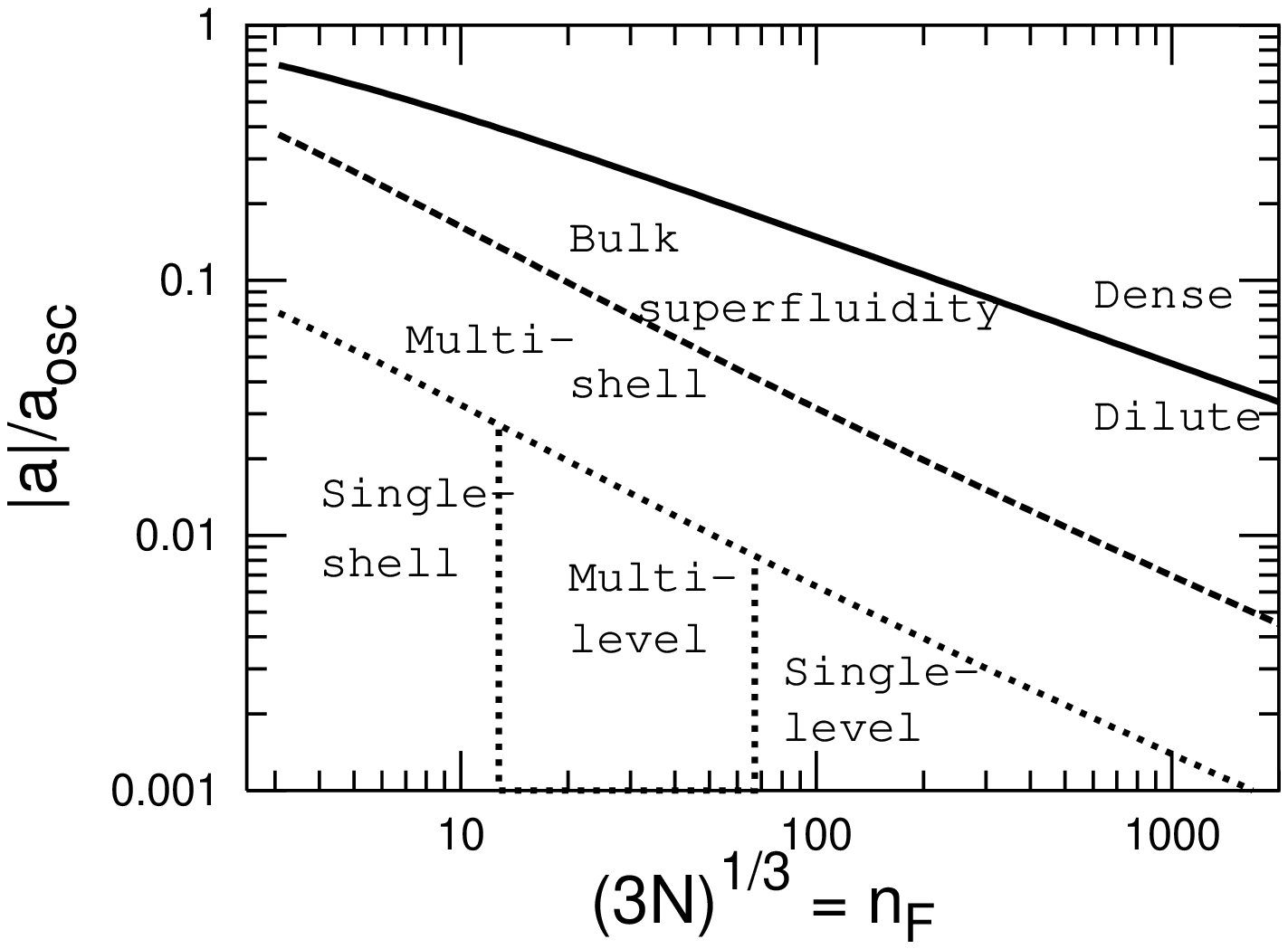,height=8.0cm,angle=-90}
\vspace{.2cm}
\begin{caption}
{Diagram displaying the regimes for the various pairing mechanisms
(see text) at zero temperature
in h.o. traps vs. the number of particles $N=n_F^3/3$
and the interaction strength $a$.
The dotted lines indicate the transitions between single-shell pairing 
$\Delta=G$, multi-level Eq. (\ref{Dml}), single-level 
Eqs. (\ref{D1}-\ref{D3}), and multi-shell pairing Eq. (\ref{DG}).
At the dashed line determined by $2G\ln(\gamma n_F)=\hbar\omega$ the pairing
gap is $\Delta\simeq\hbar\omega$, and it 
marks the transition from multi-shell pairing to bulk superfluidity
Eq. (\ref{Gorkov}). The pairing gap is $\Delta=0.54E_F$
above the full line $\rho|a|^3\ge 1$, which separates the dilute from 
the dense gas.
}
\end{caption}
\end{center}
\label{phase}
\end{figure}

\section{Pairing}

Pairing in small systems as nuclei \cite{BH}, metal clusters
\cite{metalc} and superconducting grains \cite{grains} appears in
odd-even staggering of binding energies, i.e. that systems with even
number of particles are more strongly bound than odd ones. In nuclei
the pairing is also responsible for superfluid effects in collective
motion \cite{BM}.

Pairing in finite systems is described by the Bogoliubov-de Gennes equations
\cite{deGennes} and take place between time-reversed states. As shown in
\cite{BH} these states can be approximated by h.o. wave functions in dilute
h.o. traps as long as the gap does not exceed the oscillator energy,
$\Delta\la\hbar\omega$. 
The gap equation at zero temperature then reduces to
\begin{equation}\label{gap}
 \Delta_{nl}=\frac{g}{2}\sum^{n'\la 2n_F}_{n',l'}\frac{2l'+1}{4\pi}
 \frac{\Delta_{n'l'}}{E_{n'l'}}\int_0^\infty dr r^2
 {\cal R}_{nl}^2(r){\cal R}_{n'l'}^2(r)  \,,
\end{equation}
with quasi-particle energies
\bea
  E_{n'l'} = \sqrt{(\epsilon_{n'l'}-\mu)^2+\Delta_{n'l'}^2} \,.
\eea
The cut-off $n\la 2n_F$ in the sum of the gap equation models as a
first approximation the more rigorous regularization procedure
described in Ref.\ \cite{BruunBCS} that is required for a 
delta-function pseudo-potential.

The chemical potential $\mu$ is related to the number of particles by
\bea
  N = 2\sum_{n'l'}(2l'+1)\left[1-\frac{\epsilon_{n'l'}-\mu}{E_{n'l'}}\right]\,.
\eea
For weak interactions the chemical potential lies within $ \Delta_{nl}$ from
the level energy at the Fermi surface $\epsilon_{nl}$,
except for closed levels where the chemical potential
lies between the closed and next open level.
For closed shells
$N=(n_F+1)(n_F+2)(n_F+3)/3$, the gap vanishes for interactions below a critical
value, $G\le G_c=\hbar\omega/2\ln(4\gamma n_F)$. \cite{BH}

The gap at the Fermi surface is most important and it is
generally proportional to 
the critical temperature as discussed in the
appendix. For half filled level (or shell), where
$\mu=\epsilon_{n_Fl}$, the gap is equal to the quasi-particle energy
$E_{n_Fl}=\Delta_{n_Fl}$.  We will in the following refer to this gap
as $\Delta\equiv E_{n_Fl}=\Delta_{n_Fl}$.

The h.o. traps provide a physical system in which the connection
between these methods of pairing calculations can be demonstrated and
analytical results can be given in various limits of interaction
strength and particle number.  The associated pairing regimes are
displayed in Fig. 1 and will be discussed in the following
subsections.  The first three have been discussed in earlier
publications \cite{HM,BH} but contains additional details of
derivations and sets
the notation. In the remaining three subsections, which describe new
pairing regimes, pairing gaps are calculated and the transitions to
the other regimes are discussed. The last subsection treat the
important regime of multi-level pairing which will be shown to be
especially relevant for nuclei in the following section.

\subsection{Single-level pairing}

For very weak interactions pairing take place only between 
time-reversed states $\psi_{lm}({\bf r}_1)\psi_{l-m}({\bf r}_2)$ within
the $l$-level at the Fermi surface, i.e. $l'=l$.
Maximal pairing between two particles in the level is achieved with the 
Cooper pair wave function
\bea
   \phi_{0}(n_F,l) = \sum_m \av{lml-m|00} 
 \psi_{lm}({\bf r}_1)\psi_{l-m}({\bf r}_2) \label{wf}\,.
\eea
The pairing energy
between only two particles in the level  is twice the quasi-particle energy, and so 
\bea
  E_{n_Fl} &=&\frac{g}{2}\langle\phi_{0}(n_F,l)|\delta^3({\bf r}_1-{\bf r}_2 ) 
          |\phi_{0}(n_F,l)\rangle \nonumber\\
           &=& \frac{g}{2}\frac{2l+1}{4\pi} \int_0^\infty dr r^2 \,
 {\cal R}_{n_Fl}^4(r)  \, .  \label{D1}
\eea

It follows from the seniority scheme (see, e.g., \cite{FW}) 
that this is generally the  quasi-particle energy and excitation energy 
for any even number of particles in the level.
Solving the gap Eq. (\ref{gap}) also gives $E_{n_Fl}=\Delta$ for any number
of particles in the level.

The pairing energies were given in Ref. \cite{HM}. For the top
levels $l\simeq n_F$ the h.o. wave functions are sufficiently simple
that the pairing energy can be calculated exactly. For example, ${\cal
R}_{n_Fl=n_F}^2=(2/\Gamma(l+3/2))r^{2l}\exp(-r^2)$, (in units where 
$a_{osc}=1$),  which inserted in Eq. (\ref{D1}) gives the gap
\bea \label{GnF}
   E_{n_Fl=n_F} =\frac{|a|}{\sqrt{\pi}a_{osc}}\hbar\omega \,.
\eea
The h.o. wave function for $l=0$ can be approximated by the WKB one of 
Eq. (\ref{RWKB}) when $n_F\gg 1$. 
Inserting in Eq.(\ref{D1}) gives the pair gap \cite{WKB}
\bea
   E_{n_Fl=0} =\frac{\sqrt{2}}{\pi\sqrt{n_F}}
    \frac{|a|}{a_{osc}}\hbar\omega  \,. \label{D3}
\eea
Around midshell, i.e. away from closed ($ 1\ll l$) and open ($1\ll n_F-l$)
shells, the WKB wave functions 
must be replaced near their turning points by Airy
functions which removes a logarithmic singularity and in turn 
leads to a logarithm $\ln(l)$. We find \cite{speciale}
\bea \label{Gll}
   E_{n_Fl} = \frac{\sqrt{2}}{\pi^2\sqrt{n_F}} \ln(l)
         \frac{|a|}{a_{osc}} \hbar\omega \,,\quad\quad 1\ll l\ll n_F \,,
\eea
to leading logarithmic accuracy.

As shown in Fig. 1 the single-level pairing regime appears for
sufficiently weak interactions and large number of particles 
($n_F\ga 50$) such that $D\gg G$. The corresponding pairing gaps can be
seen in Fig. 2 at large $n_F$.

\subsection{Single-shell pairing}

When $n_F$ is sufficiently small the mean-field
splitting $D$ is smaller than the pairing gaps and the
pairing acts between all states in an oscillator shell and not just in a
single $l$ multiplet, i.e. the full $SU(3)$ symmetry is effectively
restored as compared to the $SU(2)$ symmetry of the single $l$
multiplet. This enhanced pairing is referred to as
``$SU(3)$-pairing'' or ``super-pairing''. \cite{HM}
We shall calculate the gap by writing down the ``super''-pair
wave function. As in single-level pairing, however, we
obtain the same result by solving the gap equation.

Assuming $SU(3)$ symmetry the
pairing can be calculated variationally
with a  pair wave function that is a generalization of  Eq. (\ref{wf}) 
to a sum over $l=n_F,n_F-2,....,1$ or 0
\bea
   \Phi_0(n_F) = 
    \frac{\sum_l \alpha_l\, \phi_0(n_F,l)}
     {[\sum_l \alpha_l^2]^{1/2}} \,.
\eea
The weights $\alpha_l$ can be found by  a variational method
with the overlap integrals
$\int{\cal R}_{nl}^2{\cal R}_{n'l'}^2r^2dr$ calculated numerically.
\cite{speciale}
For large $n_F$ we find $\alpha_l\sim\sqrt{2l+1}$ very accurately.
The sums then reduce to the h.o. level density
\bea
 \sum_{lm}|\psi_{lm}({\bf r})|^2 =\frac{1}{4\pi} \sum_l (2l+1){\cal R}_{nl}^2 
   =\frac{\partial\rho_\sigma(r)}{\partial n}  \,. \nonumber
\eea
Here,  $\rho_\sigma=\rho/2$ is the density of one
spin state (see Eq. (\ref{rho}) for $n=n_F$)
when shells up to energy $(n+3/2)\hbar\omega$ 
are occupied and $\partial\rho(r)/\partial n$ is then the density of the $n$'th
shell.

The sum over l-states in the gap equation simplifies enormously by using
the Thomas-Fermi identity valid for  $n\gg 1$
\begin{eqnarray} \label{Trick}
   \sum_l\frac{(2l+1)}{4\pi}{\mathcal{R}}_{nl}^2(r) =
   \frac{\partial\rho_\sigma(r)}{\partial n}=
   \frac{\sqrt{2n+3}}{2\pi^2a_{osc}^3}\sqrt{1-\frac{r^2}{R^2_{TF}}} .
\end{eqnarray}
The pairing energy is thus
\bea \label{Gdef}
   E_{n_Fl} &=& \frac{g}{2}\langle \Phi_0(n_F)|\delta^3({\bf r}_1-{\bf r}_2)
             |\Phi_0(n_F)\rangle  \nonumber\\
    &=&\frac{g}{2}\frac{\int dr r^2
    [\partial\rho_\sigma(r)/\partial n|_{n_F}]^2}
   {\int dr r^2\partial\rho_\sigma(r)/\partial n|_{n_F}} \, =\, G  \,,
\eea
where the supergap $G$ was given in Eq. (\ref{G}).
The condition for single-shell pairing in atomic h.o. traps is: 
$D\la 2\Delta=2G$, is thus
$n_F\la 64/5$ according to Eqs. (\ref{G}) and (\ref{D}) or, equivalently,
to $N\la 10^3$ trapped particles.

For more than two particles in the shell we may again invoke the
seniority scheme.  It applies approximately to many superpairs when
the single-level gap is replaced by the supergap $G$, and the level
degeneracy for a full h.o. shell $\Omega=(n_F+1)(n_F+2)$. Numerical
calculations for half filled shells confirms that the pairing energy
per particle is given by the supergap for single-shell pairing
\cite{BH}.

The single-shell pair energy spectrum does, however, differ from that for
single-level pairing. In the latter case, one can generally write the pairing
wave function in Eq. (\ref{wf}) as any linear combination of the
time-reversed two-particle states $\psi_{lm}({\bf r}_1)\psi_{l-m}({\bf
r}_2)$.  Due to rotational $SO(3)$ symmetry the overlap integrals in a
single-level are independent of $m$ and the spectrum of pairing
energies is particularly simple: one state with $2E_{n_Fl}$ as given by
Eqs.(\ref{GnF}-\ref{Gll})  and
rest of the $(2l+1)$ states has zero energy per pair. In the
shell, however, the overlap between the radial wave functions depend
on $l$ and $l'$ the corresponding matrix for $l,l'=n_F,n_F-2,...,1$ or
0 has a non-trivial eigen-value spectrum: $E_S\simeq 2G/(2S+1)$,
$S=0,1,...,Int((n_F+1)/2)$, when $n_F\gg 1$ \cite{speciale}.
The corresponding eigen-vectors
$\alpha_l$ are very well approximated by Chebyshev's polynomia as function of
$x=(l/n_F)^2$ in the large $n_F$ limit.
Consequently, the excitation energy of a pair is only 2/3 of the energy $2G$
that it takes to break the pair completely.
The richer spectrum is important for expressing an effective
Hamiltonian for BCS pairing.

\subsection{Multi-shell pairing}

For increasing interaction strength pairing also take place between different
shells, i.e. $n\ne n_F$ contributes in the sum over $n$ in the gap 
equation (\ref{gap}).
The pairing strength $\Delta_n$ depends only 
weakly on $n$ for the shells around $n=n_F$ which 
mainly contribute to the pairing and we therefore approximate 
$\Delta_n\simeq\Delta_{n_F}$.
For half filled shell $\mu=\epsilon_{n_Fl}$ and $\Delta_{n_Fl}=E_{N_Fl}=\Delta$.
The gap equation then becomes \cite{BH}
\bea\label{GG}
   \Delta &=& G+ G
              \sum_{n\neq n_F}^{n\la 2n_F}\frac{\Delta}
              {\sqrt{[(n-n_F)\hbar\omega]^2+\Delta^2}}  \,.
\eea
The sum can be converted into an integral when correcting the lower limit
$n=1$ by $\gamma=e^C$ where $C=0.577..$ is Euler's constant. It then gives to
leading orders
\bea
  \Delta &\simeq& G +2G\frac{\Delta}{\hbar\omega} 
  \ln\left(\frac{2\gamma n_F}{1+\sqrt{1+(\gamma\Delta/\hbar\omega)^2}}\right) 
    \,, \label{GG2}
\eea
which for $\Delta\ll \hbar\omega$ gives
\bea\label{DG}
    \Delta &=& \frac{G}{1-2\ln(\gamma n_F)\frac{G}{\hbar\omega}} \,.
\eea
This is the  {\em multi-shell pairing} gap valid 
for interactions such that $G\ln(n_F)/\hbar\omega\la 1/2$.
 The region of multi-shell pairing
$0.1\la G\ln(n_F)/\hbar\omega\la 1/2$ extends the single-shell
pairing gap $G$ up to stronger interactions where  
$\Delta\sim\hbar\omega$ (see Fig. 1).
Cooper pairs are still essentially only formed between states 
within the same shell. However, there is pairing in many shells besides
that at the Fermi level
resulting in a gap that is larger than $\Delta=G$.

\subsection{Bulk superfluidity}

For a large system the local superfluid field is related to that
in a uniform system.
In the TF approximation the local pairing field then depends on radius
\bea
   \Delta(r) = \kappa \,  \frac{k_F^2(r)}{2m} 
   \exp\left(\frac{\pi}{2ak_F(r)}\right) \, ,  \label{Gorkov}
\eea
where $k_F(r)=(3\pi^2\rho(r))^{1/3}=\sqrt{2n_F-r^2/a_{osc}^2}/a_{osc}$ 
is the TF wave number. The prefactor is
$\kappa=8/e^{2}$ without and 
$\kappa=(2/e)^{7/3}$ with corrections from induced
interactions \cite{gap}.

When the attractions are sufficiently strong and/or the number of
particles in the trap large enough such that $G\ln(\gamma
n_F)\ga\hbar\omega$, the pairing field exceeds the h.o. frequency in
the bulk of the trap except in a narrow surface region.  For that
reason the low lying single particle excitations \cite{Baranov,BH} and
the collective modes are surface modes with typical excitation
energies of order $\sim\hbar\omega$ \cite{Bruun}.  For example, the
lowest s-wave excitation (the monopole) is $\omega_0=2\omega$ in the
collisionless limit. This is contrast to the weakly interacting
system, i.e. below the bulk
superfluidity region of Fig.1, where $\omega_0=\Delta$ for open
shells.  We refer to Ref. \cite{BH,Bruun} for a discussion of
such modes.

Here, we concentrate on the bulk superfluid field inside the trap and
will relate it to that in a uniform system.  Contrary to the surface
states, the excitations at higher energies are less affected by the
pairing field and the Bogoliubov wave functions are approximately
given by the unperturbed h.o. wave functions. Consequently, the
intra-shell paring condition discussed in Ref. \cite{BH}, which leads
to the gap equations (\ref{gap}) and (\ref{GG}), are approximately
valid for these states. However, the gap is to be
understood as an average superfluid gap $\langle\Delta\rangle$ in bulk.  The
gap equation (\ref{GG2}) is readily solved when 
$\langle\Delta\rangle\ga\hbar\omega$
\bea
  \langle\Delta\rangle \simeq 
   2n_F\hbar\omega\exp(-\hbar\omega/2G) \,. \label{expG}
\eea
We now observe that the exponent in Eq. (\ref{expG}) coincides
with that in the TF-BCS Eq. (\ref{Gorkov})
when the Fermi wave number is replaced by its spatial
average in a finite system: 
\bea
  \av{k_F(r)}\equiv \frac{\int k_F^2(r)d^3r}{\int k_F(r)d^3r}=\frac{32}{15\pi} 
                  k_F(0) \,,
\eea
where $k_F(0)=\sqrt{2n_F}/a_{osc}$.
That the spatial average should be performed over $k_F(r)^2$ follows from Eq.
(\ref{G}) where the factor $\partial\rho(r)/\partial n\propto k_F(r)$
of Eq. (\ref{Trick}) enters twice.
By definition of the supergap we find that: 
$\hbar\omega/G=\pi/2|a|\av{k_F(r)}$, and the two exponents match.
The prefactors differ
due to the approximate cutoff $n\la 2n_F$ in the gap equation, 
the approximate treatment of
the of the overlap integrals and their $n$-dependence
in the gap equation leading to Eq. (\ref{expG}). 

\subsection{Dense liquid}

The {\it dense} or strongly interacting limit, $k_F|a|\ga 1$, can be 
encountered near a Feshbach resonance where $a\to -\infty$. In this regime
the dilute gas approximation
implicit in the interactions in the Hamiltonian (\ref{H}) is no longer
valid and a new scaling region appears \cite{Thomas,Jin,Paris,HH,Carlson}. 
Both the energy per particle and the pairing gap approach a finite fraction
of the Fermi energy. Recent Greens function Monte Carlo calculations
\cite{Carlson} find that the interaction energy is -0.56 times the kinetic 
energy and that the odd-even staggering energy or pairing gap in bulk is
$\Delta\simeq 0.54E_F$. 

\subsection{Multi-level pairing}

We finally address the intermediate regime, $G\ga D\ga\Delta$,
(see Fig. 1), which lies between the single-level, single-shell and
multi-shell pairing regions studied above. Here, pairing takes
place between multiple $(l,l')$ levels lying close to the Fermi
surface $\mu\simeq\varepsilon_{n_F,l}$, and will therefore be referred as
{\em multi-level pairing}. This region overlaps into the regions studied
in the above subsections and therefore applies to most systems of trapped
fermions except traps with very few
(single-shell), very many (single-level) or strongly interacting
fermions (multi-shell).  As we shall see below nuclei can also be considered to
belong to the multi-level regime. 

For large level splitting $D\gg\Delta$ the detailed structure of the
overlap integrals in Eq. (\ref{gap}) becomes important. We shall
calculate these analytically using WKB as well as numerically from the
exact h.o.  radial wave functions. Inserting the latter in the gap equation
we calculate the pairing gaps as shown in Figs. 2+3 for half filled levels,
i.e., for $\mu=\varepsilon_{n_Fl}$, where $E_{n_Fl}=\Delta_{n_Fl}=\Delta$.

\vspace{-0.5cm}
\begin{figure}
\begin{center}
\psfig{file=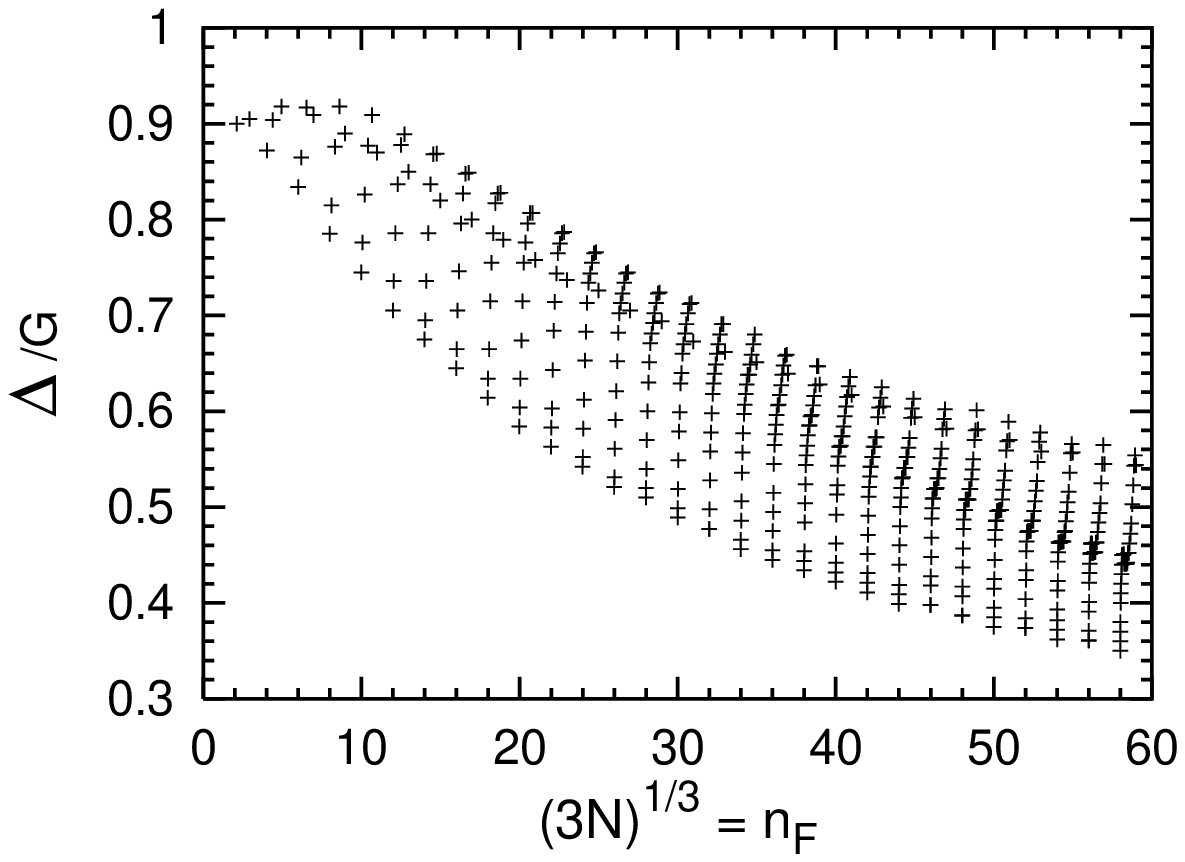,height=8.0cm,angle=-90}
\vspace{.2cm}
\begin{caption}
{The pairing gap $\Delta$ in units of $G$ vs. the number of particles 
(or $n_F$) in a trap with half filled levels 
calculated from the gap equation (\ref{gap}).
The scattering length is $a=-0.01a_{osc}$.
The pairing gaps $\Delta_{n_Fl}$ are plotted with + for each half filled
level $l$ (only even $l$ and $n_F$ are shown).
The gap undergoes a transition
from single-shell to multi-level and finally single-level pairing.
}
\end{caption}
\end{center}
\label{a3}
\end{figure}

A qualitative understanding of these results can be obtained from
analytical WKB calculations.
At large $n_F$ the WKB wave functions ${\cal R}_{n_Fl}$ of Eq. (\ref{RWKB})
are excellent approximations 
except for the multi-level overlap, $l=l'$ where a logarithmic singularity 
appear, which was responsible for the $\ln(l)$ in Eq. (\ref{Gll}).
Otherwise,  when $0<|l^2-l'^2|\ll n_F^2$
the overlap integral of the WKB wave functions of
Eq. (\ref{RWKB}) is dominated by the overlap near the
turning points. Defining: $\xi=2n_F(r/a_{osc})^2-(r/a_{osc})^4$, the overlap is
\bea
\int_0^\infty dr r^2
   {\cal R}_{n_Fl}^2(r){\cal R}_{n_Fl'}^2(r) &\propto &
  (\frac{4}{\pi})^2 \int \frac{dr}{\sqrt{\xi-l^2}\sqrt{\xi-l'^2}} \nonumber \\
       &\propto &    - \ln(|x-x'|) \,, \label{log}
\eea 
to leading logarithmic order, where $x=(l/n_F)^2$ and $x'=(l'/n_F)^2$.
This logarithmic dependence of the matrix element has the interesting 
consequence that it leads to a {\it double log} in the gap equation. 
When the energy factor 
$1/E_{nl'}\simeq(D^2(x-x')^2+\Delta^2)^{-1/2}$ is summed over
$l'$ or, equivalently when $n_F\gg 1$, integrated over $x'$, 
it leads to the (usual) factor $\ln(D/\Delta)$. The other logarithm appears 
from the matrix element of Eq. (\ref{log})
which, when summed over $l'$ in the gap equation with the energy factor, attains
the lower cutoff at $|x-x'|\simeq \Delta/D$.
It is assumed that the $l$-dependence of the gap is sufficiently weak within
the levels over which pairing take place, i.e. $\Delta\ll D$, which is
supported by numerical results
(see Figs. 2+3).
The gap equation then reduces to:
\bea
  1 &\simeq& \alpha(x) G\int_0^1 dx' 
   \frac{-\ln|x-x'|}{\sqrt{D^2|x-x'|^2+\Delta^2}}
   \nonumber \\
    &\simeq&   \alpha(x) \frac{G}{D} \ln^2(\beta(x) D/\Delta) \,,
\eea
to leading logarithmic orders.
The factors $\beta(x)$ and $\alpha(x)$ both
depend on $x=(l/n_F)^2$, but not on $l$ or $n_F$ separately.
Only $G$ and $D$ depend on $n_F$ explicitly.

The resulting multi-level gap is
\bea  \label{Dml}
   \Delta = \beta(x)\, D \exp\left[ -\sqrt{D/G\alpha(x)} \right] \,.
\eea
At mid-shell, $x=1/2$, the exponent
can be calculated: $\alpha(1/2)\simeq15/32\sqrt{2}$, and the prefactor:
$\beta(1/2)=2e$, to leading logarithmic order for large $n_F$.
Near open ($l=0$) and  closed ($l=n_F$) shells
there are generally only half as many states to
pair with which reduces $\alpha(x=0)$ and $\alpha(x=1)$
by a factor $\sim 1/2$ on
average. However, as the matrix elements are larger for  $l\simeq n_F$ but
smaller for $l=0$ (see Eq.(\ref{log})), the gaps become asymmetric
with a maximum above midshell. Such a shell structure is also found 
in the multi-level pairing regime when
the gap equation is solved numerically with the exact h.o. wave functions
as is shown in Figs.(2+3). 

Two illustrative examples are shown in Figs. (2+3). In the first the
coupling is sufficiently weak, $a=-0.01a_{osc}$, that the pairing
undergoes transitions from from single-shell at small $n_F$ to
multi-level for $10\la n_F\la 40$ and finally single-level pairing for
large $n_F$.  In Fig. 3 the coupling is stronger, $a=-0.04a_{osc}$,
and the pairing undergoes transitions from from single-shell to
multi-level and approaches multi-shell pairing for large $n_F$. These
transitions between pairing regimes are illustrated in Fig. 1.

\vspace{-0.5cm}
\begin{figure}
\begin{center}
\psfig{file=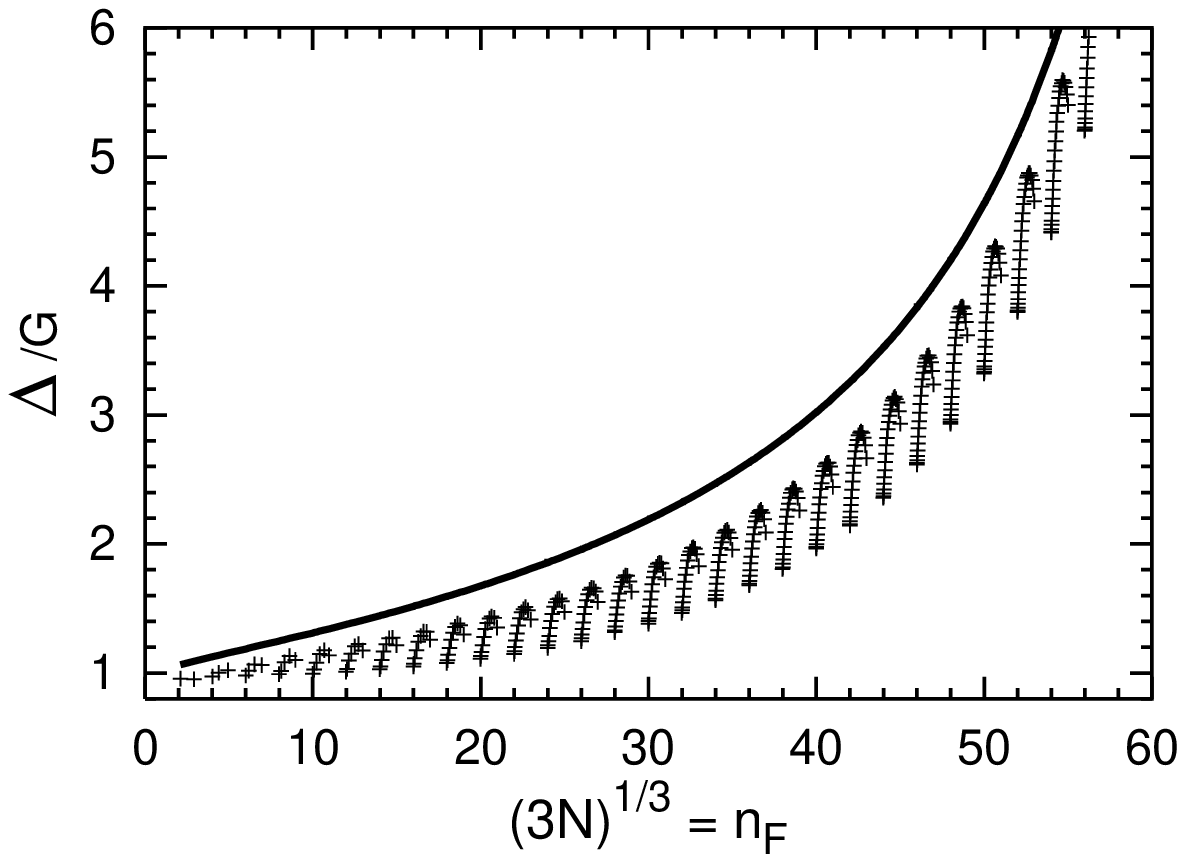,height=8.0cm,angle=-90}
\vspace{.2cm}
\begin{caption}
{Same as Fig. 2. but with $a=-0.04a_{osc}$. Full line is the multi-shell gap
$G/[1-2\ln(\gamma n_F)G/\hbar\omega]$.
The gap undergoes a transition from single-shell to a mixture of 
multi-level and  multi-shell pairing.}
\end{caption}
\end{center}
\label{a1}
\end{figure}

The multi-level gap of Eq. (\ref{Dml}) is quite
robust and applies to many systems as long as $G\ga D\ga\Delta$. Its validity 
for nuclei will be discussed below.
The formula for the multi-level gap can be generalized by relating the shell
splitting to the density of states at the Fermi surface as
$\partial n/\partial\epsilon =\Omega/D$, where $\Omega=n_F^2$ is the
number of states in the shell. Likewise, the supergap is related to
the coupling constant as $G\propto g\Omega$. If the level spectrum of
Eq. (\ref{EMF}) is changed, the multi-level gap remains valid,
if the shell splitting is correspondingly scaled with the level
density.  In other words the double log does not depend on the details
of the level spectrum but is generic for systems like
h.o. traps, because the one log is associated with the overlap
matrix elements between nearby states 
and the other is associated with the quasi-particle energy in the gap
equation.

The appearance of a double log is not unique for multi-level pairing
but also occurs in the case of color superconductivity in quark matter
\cite{Son}.  The physical mechanism behind is, however,
different. Within perturbative QCD the singular quark-quark
interaction $g_{QCD}/q^2$, where $q$ is the momentum transfer carried
by a gluon, is dynamically screened by Landau damping. When the
interaction is integrated over momentum and energy transfer, the
dynamical screening leads to a logarithm of the gap. Thus it is the
interactions that are responsible for the second logarithm and not the
wave function overlap of nearby states as in the h.o. trap.

\section{Pairing in nuclei}

The nuclear mean field is often approximated by a simple h.o. form and
the residual effective pairing interaction by a delta force in order
to obtain some qualitative insight into single particle levels,
pairing, collective motion, etc., (see, e.g., \cite{BM,FW}).  We can
therefore compare pairing in nuclei to that in traps as
investigated above, once the h.o. potential is adjusted to describe nuclei. We
emphasize that we do not intend to calculate the quantitative pairing
gaps for each individual nucleus which would require detailed
knowledge of the individual level spectra, deformation, many-body
effects, etc.  Instead we aim at qualitative results for the
pairing gap dependence on mass number, shell
effects, and to extrapolate to very large nuclei, nuclear and neutron
star matter.

Large nuclei have approximately
constant central density $\rho_0\simeq 0.14$~fm$^{-3}$ and Fermi energy
$E_F$ in bulk. Therefore the
h.o. frequency, which is fitted to the nuclear mean field, 
decreases with the number of nucleons $A=N+Z$, where $N$ now
is the number of neutrons and $Z$ the number of protons in the nucleus, as
\bea
 \hbar\omega\simeq E_F/n_F \simeq 41 {\rm MeV}\cdot A^{-1/3} \,. \label{omega}
\eea
In the valley of $\beta$-stability the number of protons is
$Z\simeq A/(2+0.0155A^{2/3})$.

Secondly, the nuclear mean field deviates from a h.o. potential by being 
almost constant inside the nucleus and vanish outside. The resulting net
anharmonic nuclear field is {\it stronger} and {\it opposite} in sign to the
corresponding (anharmonic) mean field in atomic traps. Therefore, the level
splitting is larger and the ordering
of the l-levels is reversed. In addition,
a strong spin-orbit force splits the single particle
states of total angular momentum $j=l\pm 1/2$, such that the
$j=n_F+1/2$ is lowered down to the shell ($n_F-1$) below.

Proton and neutron pairing gaps are typical of order $\sim 1$~MeV in
nuclei, i.e., smaller than both $\hbar\omega$ and $D$ but of order the
average splitting between two adjacent j-levels.  Consequently, nuclei
can be considered as h.o. traps with a level splitting $D$ such that
they fall into the multi-level pairing regime. However, the number of
particles in nuclei is relatively small that nuclei are close to the
single shell pairing regime. Furthermore, the interactions are so
strong that multi-shell pairing also becomes important. From the results
of Sec. III we can predict several features of neutron and proton
pairing gaps:

\begin{itemize}

\item {\bf Mass scaling:} 
Since $\hbar\omega$ scale as $\sim A^{-1/3}$
and $a_{osc}\propto n_F^{1/2}$ (see Eq. (\ref{G})) the single-shell
pairing gap also scales as $G\sim A^{-1/3}$.  The level splitting tend
to reduce the pairing towards the multi-level gap but is
compensated by multi-shell pairing. Therefore, the pairing gaps in light and
medium mass nuclei scale approximately as:
$\Delta\simeq G \propto A^{-1/3}$.

\item {\bf Shell structure:} The pairing gaps should exhibit a strong
shell structure similar to those for multi-level pairing (Figs. 2+3),
however with reversed $l$ since $D$ is negative for nuclei. Due to the
strong spin-orbit force the $j=l\pm 1/2$ states are split and the
$j=n_F+1/2$ is lowered down to the shell below.  The magic numbers become
$N,Z=8,14,28,50,82,126,184,..$, etc.  rather than the h.o. filled
shell particle numbers $N,Z=2,8,20,40,70,112,168,240,...$, etc.

\item {\bf Bulk limit for large nuclei/nuclear matter:} 
For very large
nuclei multi-shell pairing becomes important and pairing
approaches that in bulk matter.
By fitting the effective nucleon coupling constant to pairing
gaps in finite nuclei we will below estimate the pairing in nuclear
matter from Eq. (\ref{Gorkov}).

\end{itemize}
These predictions agree qualitatively with experimental data (see Figs. 4+5).

For a more quantitative calculation the level spectrum must be specified.
Instead of fitting the level spectrum for each nucleus with a correspondingly
large number of adjustable parameters, we make the following 
simplifying approximation analogous to the single particle spectrum of
Eq. (\ref{EMF})
\bea
   \epsilon_{nj} = \left( n+\frac{3}{2} \right) \hbar\omega 
           \,-\, D \frac{j(j+1)}{(n+1/2)(n+3/2)} \,. \label{ls}
\eea
The level splitting is approximated by 
$D=0.13(n_F+3/2)\hbar\omega$ in analogy with Eq. (\ref{D}) and based on
nuclear structure calculations, which generally find 
that the level splitting increase with shell number and is of order 
$D\simeq\hbar\omega$ for heavy nuclei ($n_F\sim6$). The spin-orbit splitting
and resulting change in magic numbers are incorporated to first approximation 
by simply moving the lowest level in each
shell down to the shell below, i.e., $D=\hbar\omega$ for $j=n_F+1/2$.

The pairing gaps and quasi-particle energies
can now be calculated by solving the gap equation inserting the
h.o. matrix elements and with $\hbar\omega$ and level splitting as
given by Eqs. (\ref{omega}) and (\ref{ls}).
The effective strength $a$ is then the only adjustable parameter.

The data on neutron and proton pairing
is obtained from the odd-even staggering of nuclear
binding energies $B(N,Z)$. It has been shown that mean field contributions
can be removed \cite{Naza} by using the three-point filter 
\bea
   \Delta^{(3)}(N) &\equiv& \frac{(-1)^N}{2} [B(N-1,Z)+B(N+1,Z) \nonumber\\
      &&  \quad   -2B(N,Z)] \,,
\eea
when $N$ is an odd number of neutrons. The analogous relation is valid 
for protons.

The total binding energies are elaborate sums over quasi-particle
energies weighted with occupation numbers. Mean field energies and
single particle energies should also be included self
consistently. Furthermore, the Bogoliubov transformation does not
preserve exact particle number and does not treat a single un-paired
particle that appears in odd-number systems.  Also nuclear spectra are
complicated by deformations and the finite range, spin and spin-orbit
dependences of the nucleon-nucleon interaction \cite{BM,RS}. For these
reasons we cannot calculate $\Delta^{(3)}$ directly. Instead we will
for simplicity compare to the quasi-particle energies and pairing
gaps, which are calculated directly from the gap equation as function
of $\mu(N,Z)$. It has been argued (see, e.g.  \cite{Naza} and
refs. herein) that mean-field effects cancel in the quasi-particle
energy for odd particle number and that it therefore may be compared
to the corresponding three-point energies. However, as particle number
fluctuations, deformations, possible mean field energy corrections and
other effects are not included, the pairing gaps are also shown in
Figs. (4+5) for comparison. They are equal to the quasi-particle
energy for half-filled levels only and generally smaller especially 
at the magic numbers where the gaps vanish in several cases.

We compare in Fig. 4 to the experimental
$\Delta^{(3)}(N)$ averaged over isotopes with the calculated gaps
$\Delta_{n_Fl}$, and in Fig. 5 the analogous for protons
$\Delta^{(3)}(Z)$ averaged over isotones. In the calculations the
effective coupling is the only parameter fitted to experimental data. 
For both neutrons and protons we extract $a\simeq -0.41$~fm.

\vspace{-0.5cm}
\begin{figure}
\begin{center}
\psfig{file=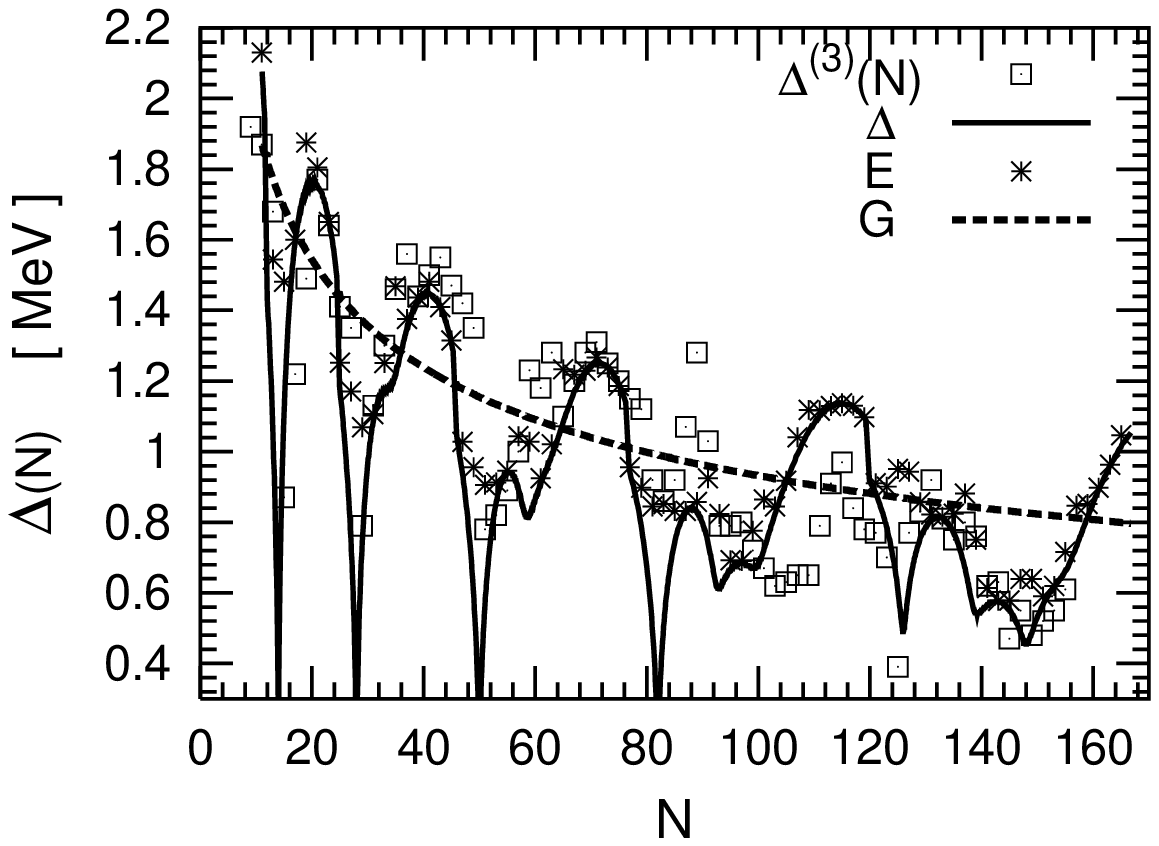,height=8.0cm,angle=-90}
\vspace{.2cm}
\begin{caption}
{Neutron pairing energies
vs. the number of neutrons. The experimental odd-even staggering energies
$\Delta^{(3)}(N)$ are averaged over
isotopes \protect\cite{Audi,Naza}. The calculated gaps $\Delta$ and
quasi-particle energies $E$  are obtained
from the gap equation (see text) with effective coupling strength 
$a=-0.41$~fm. The supergap $G$ is shown with dashed line.}
\end{caption}
\end{center}
\label{N}
\end{figure}
\vspace{-0.5cm}

We note that although the neutron pairing gaps in Fig. 4 generally are
larger than the proton ones in Fig. 5, this is not reflected in the
effective coupling constants. The reason is the asymmetry of heavy
nuclei. For example, for $N=82$ the mass number is $A\simeq 140$
whereas for $Z=82$ it is $A\simeq 208$. The mass numbers enter both
$D$ and $G$ and lead to a reduction of the proton pairing gap relative to
the neutron one by just the right amount so that the experimental data on
neutron and proton odd-even staggering can be fitted with the same pairing
strength $a=-0.41$~fm.

Considering the simplicity of the model it describes a large number of
experimental gaps fairly well on average.  In a number of cases,
however, the calculated pairing gaps differ significantly from the
measured neutron gaps. Some of these deviations can be attributed to
the crude single particle level spectra assumed.  If the single
particle level energies are adjusted according to more detailed mean
field calculations (see, e.g., \cite{BM}) the agreement with
experimental pairing gaps improves in several cases.

\begin{figure}
\begin{center}
\psfig{file=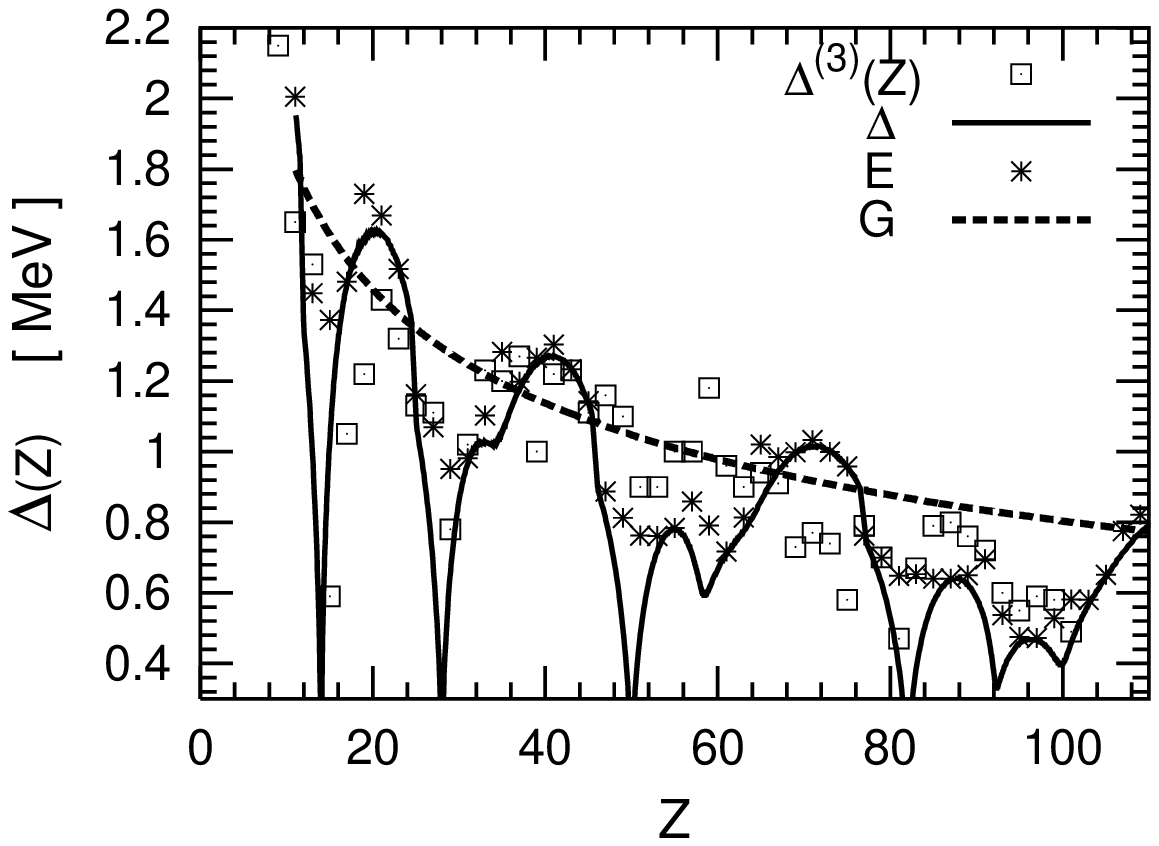,height=8.0cm,angle=-90}
\vspace{.2cm}
\begin{caption}
{As Fig. 4 but for protons}
\end{caption}
\end{center}
\label{Z}
\end{figure}

The pairing gaps are sensitive to the shell splitting and the
coupling. The uncertainty in the coupling is smaller because the
changes in $D$ affects both the exponent and the prefactor in the gap
of Eq. (\ref{Dml}) in a compensating way.  Another uncertainty arise
from the upper cutoff.  Whereas h.o. traps can pair between shells up
to $\sim2n_F$, nuclei have a continuum of states at about the binding
energy per nucleon $E_B\simeq 8$~MeV above the Fermi level.  Therefore
the sum over shells in the gap equation has been limited to
$n\la(E_F+E_B)/\hbar\omega$, which for medium mass nuclei corresponds
to $n\la n_F+1$, with a smooth cutoff as in Ref. \cite{Bender}. The
pairing gaps are, however, only logarithmically sensitive to this
cutoff.

 Nuclei can also be deformed around mid-shell which
increase pairing \cite{Naza}. These and other nuclear many-body
effects must be included in a more quantitative study of pairing in
individual nuclei. For this purpose it will be useful also to study
pairing in elongated atomic traps with deformations like in nuclei.
Near closed shells pairing reduce or inhibit deformations.
The effect of pairing is also observed in rotational spectra where the moment
of inertia is reduced from the rigid to superfluid or irrotational value
\cite{BM}. In traps the external h.o. potential generally dominates over the
mean field interaction energy and thus does not deform spontaneously 
\cite{unpublished} as in nuclei
- unless of course if the h.o. trap itself is deformed.

Nuclei cannot directly be placed in any of the various pairing phases
of Fig. 1 because the level splitting is larger than for trapped
atoms. The effective scattering length $a=-0.41$~fm and
$a_{osc}\simeq1$fm$/\sqrt{n_F}$ would place nuclei with masses up to
$A=250$ corresponding to $n_F\la6$ in the upper left corner of Fig. 1
in the multi-shell region. However, because $D$ is much larger in
nuclei than for trapped atoms, the multi-level pairing region extends
down to lower $n_F$ and up to larger strengths $|a|$.  Furthermore, the
continuum of states in nuclei reduce the effect of multi-shell pairing
as discussed above. Therefore, nuclei rather belong to the transition
region between single-shell, multi-shell and multi-level pairing.

The best fit to odd-even staggering energies of nuclei determines
$a=-0.41$~fm accurately. Systematic errors may, however, be expected
from the approximations implicit in the level splitting, the cut-off,
and in approximating $\Delta^{(3)}$ by $\Delta$. However, because the
multi-level and the multi-shell pairing partly compensate, a good
approximation to the average gap in nuclei is the single-shell
supergap
\bea
  \Delta\simeq G\simeq \frac{|a|}{0.41{\rm fm}} \frac{5.5{\rm MeV}}{A^{1/3}} 
  \,.
\eea
This supergap is also shown in Figs. (4+5). It
does not depend on the level-splitting or cut-off and is therefore a 
robust prediction for the average magnitude and mass scaling of pairing gaps
in nuclei.

Empirically the pairing term in Bethe-Weisz\" acker
liquid-drop formula, $\Delta\simeq 12$~MeV$\cdot A^{-1/2}$, 
fits the odd-even staggering energies
of nuclei with $A\la 250$ averaged over shell effects. 
The scaling with mass number can now be understood in terms of the supergap 
with shell corrections. If $D$
was a constant times $\hbar\omega$ then the multi-level gap of
Eq. (\ref{Dml}) would also scale as $A^{-1/3}$. However, because $D$
increase with $n_F\propto A^{-1/3}$, the multi-level gap decrease
faster with $A$ for small and medium mass nuclei but slower for heavy nuclei due to multi-shell pairing. Both are in accordance with
the empirical mass dependence.

The pairing in nuclear matter can also be estimated once the effective
interaction has been determined.  Inserting in Eq. (\ref{Gorkov})
$a=-0.41$~fm and $k_F=1.33$~fm$^{-1}$ at nuclear saturation density,
$\rho_0=0.15$~fm$^{-1}$, we obtain the proton and neutron pairing gaps
\bea
  \Delta \simeq 1.1 {\rm MeV} \,, 
\eea
in the bulk of very large nuclei and in symmetric nuclear matter at
nuclear saturation density.  This number is compatible with earlier
calculations \cite{BP79} of the $^1S_0$ pairing gap in nuclear and
neutron star matter around normal nuclear matter densities.
A smaller value for bulk pairing might have been expected from Figs. (4+5) by
extrapolating the odd-even staggering energies of heavy nuclei to
higher mass numbers. However, as $A\propto n_F^3$ becomes large the
multi-shell pairing contribute with the increasing term $\log(n_F)$ 
in the gap equation. Therefore the bulk
value is larger than the pairing gap in heavy nuclei which also have a
smaller cutoff due to continuum states as discussed above.

Neutron star matter has a wide range of densities and is very
asymmetric, $Z/A\simeq 0.1$. One can attempt to estimate of the pairing gaps as
function of density from the gap in bulk, Eq. (\ref{gap}), with
$a\simeq -0.41$~fm and the neutron or proton Fermi wave numbers
$k_F^{N,Z}=(3\pi^2\rho_{N,Z})^{1/3}$ as function of densities. 
However, the effective interaction $a$ is
 density dependent. At higher densities we expect the effective interaction to
become repulsive as is the case for the nuclear
mean field at a few times nuclear saturation
density. At lower densities the effective scattering length should approach
that in vacuum which for neutron-neutron scattering
is $a(^1S_0)\simeq -18$~fm. This dilute limit $k_F|a|\ll1$ does, however,
require extremely low densities as compared to normal nuclear matter density.

\section{Summary}

 Pairing gaps have been calculated for ultracold atomic Fermi gases in
harmonic oscillator traps and compared to nuclei. The pairing
mechanism was found to be similar for these systems in the sense that
the spacing between single particle states (and shells) reduce the
pairing over several of these levels near the Fermi surface referred
to as multi-level pairing.
At low particle densities the shell structures in traps
are pronounced as they are in nuclei and the level degeneracies are
important for the size of the gaps which can differ substantially from
those known from homogeneous systems \cite{gap} and systems with
continuous level densities. 

Neutron and proton pairing gaps in nuclei were calculated and with an
effective coupling strength $a=-0.41$~fm a qualitative description of
their shell structure could be given, and the average pairing gaps
were found to scale with mass number approximately as
$\Delta\simeq5.5$~MeV$/A^{1/3}$ as predicted from the supergap.
Eventually for large mass number the gap approaches the $^1S_0$
superfluid gap in uniform nuclear matter which was calculated from the
as $\Delta\simeq 1.1$~MeV for both neutrons and protons.

Mixing fermionic with bosonic atoms improve cooling \cite{Stoof,Ketterle} to
lower temperatures so that weak pairing can also be studied, and the
additional induced interactions between fermions and bosons generally
enhance pairing \cite{gap}. Furthermore, the shell splitting can be
changed in a controlled way by the number of bosons in the trap and
the sign and strength of their interaction with the fermions.

 The similarity of  multi-level and bulk pairing in atomic traps and nuclei 
may provide new insight into pairing and superfluidity in nuclei and
neutron star matter from tabletop experiments at low temperatures.

\appendix
\section{Critical temperatures}

The pairing gaps generally decrease with increasing temperature from
its zero temperature value $\Delta(T=0)$, that was calculated above,
to the critical temperature $T_c$, where it vanishes. 
The temperature dependence of the gap and $T_c$ itself
are determined from
\bea
 \Delta_{n_F,l}(T) &=& \sum_{n'l'}  [1-2f(E_{n',l'}/T)]
 \frac{\Delta_{n',l'}(T)}{E_{n',l'}} \frac{g}{4\pi}   \nonumber \\ 
 && \times  \int_0^\infty dr r^2{\cal R}_{nl}^2(r){\cal R}_{n'l'}^2(r)
    \,, \label{T}
\eea
with $E_{n'l'}=\sqrt{(\epsilon_{n'l'}-\mu)^2+\Delta^2(T)}$.

As shown in \cite{BruunT}  $T_c$ is exactly
half of the zero temperature gap
\bea \label{Tc1}
  T_c = G/2(1-2\ln(\gamma n_F) G/\hbar\omega) = \frac{1}{2}\Delta(T=0) ,
\eea
in  the single-shell and multi-shell pairing 
regimes, and this also applies to the
single-level pairing regime.

In uniform Fermi gases the critical temperature is \cite{gap}
\bea
  T_c = \frac{\gamma}{\pi} \kappa \, E_F\, \exp(2/\pi ak_F) 
      = \frac{\gamma}{\pi}\Delta(T=0) \,.
\eea
The ratio  $T_c/\Delta(T=0)=\gamma/\pi\simeq0.567$ is the same irrespective of
whether induced interactions are included or not. 

In the multi-level regime $T_c$ can be determined from
Eq. (\ref{T}) with overlap integrals as given in Eq. (\ref{log}).
To leading logarithmic order we find for large $n_F$
\bea \label{Tc3}
  T_c =  \frac{\gamma}{\pi} \Delta(T=0) \,.
\eea
That $T_c/\Delta(T=0)=\gamma/\pi$ as in the uniform Fermi gas is mainly because
the pairing take place over several l-levels and the level density 
therefore is effectively continuous. The overlap integrals do not change this
ratio to leading logarithmic order.

Although $T_c/\Delta(T=0)=\gamma/\pi\simeq0.567$ is close numerically to
the value 1/2 found in the single-shell, single-level and multi-shell
regimes, the difference reveals the qualitative differences in the
underlying level spectrum, namely, continuous vs. discrete respectively.

Near a Feshbach resonance the strongly interacting Fermi gas becomes
unstable towards molecule formation and $T_c$ for BCS superfluidity is
expected cross-over towards the slightly smaller critical temperature
for forming a Bose-Einstein condensate of molecules \cite{Randeria}.
Both the BCS and BEC critical temperatures are, however, above the
lowest temperatures achieved recently for trapped Fermi atoms
\cite{JILA,Thomas,Ketterle} if we assume $T_c\simeq 0.5\Delta$ as in
Eqs. (\ref{Tc1}-\ref{Tc3}) and take $\Delta=0.54E_F$ according to
Ref. \cite{Carlson}.


\begin{thebibliography}{99}
\bibitem{WKB} The WKB approximation
can be applied between the turning points
$r^2_\pm=[n_F+3/2\pm\sqrt{(n_F+3/2)^2-l(l+1)}]a_{osc}^2$.
Near the turning points the wave functions 
must be replaced by Airy functions for detailed gap calculations.
The phase $\theta$ is only important to leading log order when 
$l=0$ where $\theta=0$.
When the wave function has many nodes $1\ll l\ll n_F$, 
the oscillations can be averaged $\av{\sin^2(kr)}=1/2$ and 
$\av{\sin^4(kr)}=3/8$.
\bibitem{TF3} For a finite number of particles the
factors $n_F+3/2$ includes a correction to $n_F$, which 
has been checked numerically to improve the TF approximation.
\end{thebibliography}
\end{document}